\documentclass[usegraphicx]{mn2e}
\usepackage{subfigure}
\newcommand{\rsun}{$\mathrm{R_{\sun}}$}
\newcommand{\msun}{$\mathrm{M_{\sun}}$}

\newcommand{\rate}{$\mathrm{M_{\sun}} \, \mathrm{yr}^{-1}$}

\newcommand{\xpn}[2]{$#1 \times 10^{#2}$}
\newcommand{\kms}{$\mathrm{km \, s^{-1}}$}

\title[The formation of the double neutron star pulsar J0737--3039]
      {The formation of the double neutron star pulsar J0737--3039}
\author[Dewi \& van den Heuvel]
       {J. D. M. Dewi$^{1,2}$\thanks{email: jasinta@science.uva.nl}, 
        E. P. J. van den Heuvel$^1$\\
        $^1$Astronomical Institute {\it Anton Pannekoek},
            University of Amsterdam, Kruislaan 403, NL-1098 SJ Amsterdam,
            The Netherlands\\
        $^2$Bosscha Observatory and Department of Astronomy, Institut Teknologi Bandung, 
            Jl. Ganesha 10, Bandung 40132, Indonesia}
\date{Accepted . Received ; in original form }

\pagerange{\pageref{firstpage}--\pageref{lastpage}}
\pubyear{}

\begin{document}

\maketitle

\label{firstpage}

\begin{abstract}
We find that the orbital period (2.4 hours), eccentricity (0.09), dipole magnetic field strength (\xpn{6.9}{9}~Gauss) and spin period (22 ms) of the new highly relativistic double neutron star system PSR J0737--3039 can all be consistently explained if this system originated from a close helium star plus neutron star binary (HeS-NS) in which at the onset of the evolution the helium star had a mass in the range 4.0 to 6.5~\msun\ and an orbital period in the range 0.1 to 0.2 days. Such systems are the post-Common-Envelope remnants of wide Be/X-ray binaries (orbital period $\sim$ 100 to 1000 days) which consist of a normal hydrogen-rich star with a mass in the range 10 -- 20~\msun\ and a neutron star. The close HeS-NS progenitor system went through a phase of mass transfer by Roche-lobe overflow at a high rate lasting a few times $10^{4}$~years; assuming Eddington-limited disk accretion onto the neutron star this star was spun up to its present rapid spin rate. At the moment of the second supernova explosion the He star had a mass in the range 2.3 to 3.3~\msun\ and in order to obtain the present orbital parameters of PSR J0737--3039 a kick velocity in the range 70 -- 230~\kms\ must have been imparted to the second neutron star at its birth.
\end{abstract}


\section{Introduction}
\label{0737:sec:intro}

A new double neutron star (DNS) system, J0737--3039, was recently discovered 
 (Burgay et al. 2003). This is the fifth galactic DNS; all share the same characteristics of having eccentric orbits ($e \ga 0.08$) and short pulse periods (22 -- 104 ms) and low period derivatives ($\sim 10^{-18} \, \mathrm{s \, s^{-1}}$). The two latter parameters imply a weak surface magnetic field ($\sim 10^{10}$~G) which indicates that the pulsar has been recycled (Smarr \& Blandford 1976; Alpar et al. 1982; Radhakrishnan \& Srinivasan 1982), and that the observed pulsar in each system is the older neutron star and has gone through a mass-transfer phase (Srinivasan \& van den Heuvel 1982). 

Dewi \& Pols (2003) found that the wide-orbit DNSs J1518+4904 and J1811--1736 are produced from helium star-neutron star binaries that avoided Roche-lobe overflow (RLOF); while the close-orbit B1913+16 and B1534+12 can be formed either with or without a mass-transfer phase in the helium star-neutron star binary stage of evolution. Among the five galactic DNSs, J0737--3039 has the shortest spin period (22 ms), shortest orbital period ($0\fd102$), and lowest eccentricity (0.088). In this paper we examine the possible formation history of J0737--3039. We will show that unlike the other DNSs, J0737--3039 can only be formed from helium star-neutron star binaries which have gone through RLOF.

\section{The formation of J0737--3039}
\label{0737:sec:formation}

\subsection{Possible progenitors}
\label{0737:subsec:progenitor}

We examine the formation history of the system is terms of the so-called standard scenario (Bhattacharya \& van den Heuvel 1991), in which DNSs were formed from Be/X-ray binaries. When the Be star leaves the main sequence and becomes a giant, a runaway mass transfer takes place due to the large mass ratio of the system. The neutron star is engulfed by the giant's envelope, and a common-envelope (CE) phase ensues in which the neutron star spirals down to the helium core of the giant, ejecting its hydrogen-rich envelope in this process. The outcome of this spiral-in is a very close binary consisting of a helium star and a neutron star (HeS-NS). For companions less massive than about 5~\msun, this mass transfer can still be stabilized such that the formation of a CE can be avoided (e.g. King \& Ritter 1999; Podsiadlowski \& Rappaport 2000; Tauris, van den Heuvel, Savonije 2000). However, for more massive companions such as the Be stars considered here the formation of a CE cannot be avoided.

\begin{figure}
  	 \centerline{\includegraphics[width=84mm]{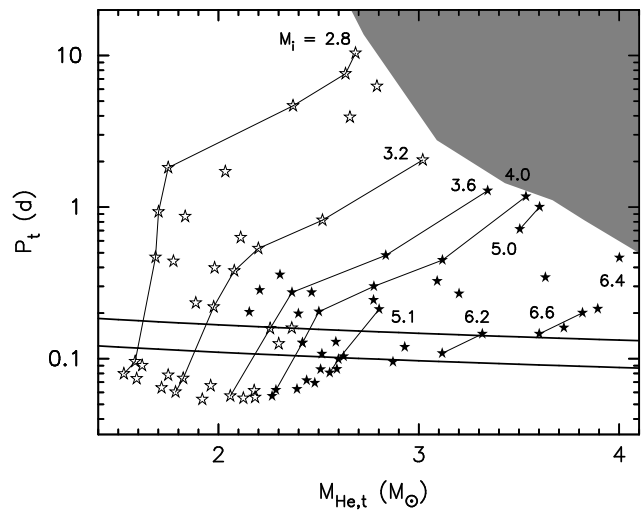}}
	 \caption[]{The helium star masses and the orbital periods after a mass-transfer
                    phase in HeS-NS binaries, taken from Dewi et al. (2002) and Dewi \&
                    Pols (2003). The open star symbols mark the systems which go through 
                    a CE phase due to the development of convective helium envelope. Thin 
                    lines connect the remnants of systems with the same initial helium 
                    star masses, $M_{\mathrm{i}}$. The shaded area marks the region where
                    the helium stars never fill their Roche lobes. The maximum and 
                    minimum pre-SN orbital period of J0737--3039 are represented by the
                    upper and lower thick lines, derived from the separation of the 
                    system at the time it was born, i.e. a characteristic spin-down
                    time ago.}
	 \label{0737:fig:presn}
	\end{figure}

The helium star evolves further through core-helium burning and further burning stages. During this evolution it transfers mass to the neutron star by RLOF and also loses mass by stellar wind. Fig.~\ref{0737:fig:presn} depicts the parameters of these HeS-NS binaries at the end of this RLOF and wind mass loss, resulting from detailed evolutionary calculations (Dewi et al. 2002; Dewi \& Pols 2003). The number indicated at each curve represents the initial mass of the helium star at the beginning of its evolution. The stars indicate the outcome of the evolution of systems with a range of initial orbital periods (note that the lines connecting the stars are not evolutionary tracks). Also presented in the figure are the maximum and minimum possible orbital periods of the progenitor of J0737--3039 just prior to the supernova (SN) explosion. These lines are derived with the assumption that the pre-SN (circular) orbital separation, $a_{\mathrm{t}}$, lies between the periastron and apastron of the post-SN separation, i.e.
	\begin{eqnarray}
	a_{\mathrm{f}} \, (1-e_{\mathrm{f}}) \leq 
	a_{\mathrm{t}} \leq 
	a_{\mathrm{f}} \, (1+e_{\mathrm{f}})
	\label{0737:eq:post-sep}	 
	\end{eqnarray}
(Flannery \& van den Heuvel 1975), where $a_{\mathrm{f}}$ and $e_{\mathrm{f}}$ are the post-SN separation and eccentricity. Here $a_{\mathrm{f}}$ and $e_{\mathrm{f}}$ are the parameters at the time when it was born, which is assumed to be a characteristic time ago, $\tau = P / 2 \dot{P}$. These parameters were derived by integrating the effects of gravitational-wave radiation on the observed ($a$, $e$) over $\tau$.

Dewi \& Pols (2003) found that systems represented with the open star symbols undergo a common-envelope (CE) phase due to the development of a helium convective envelope. The consequence of this is that, after the CE phase these systems will have a lower mass ($M_{\mathrm{He,t}} \sim 1.4$~\msun) and much shorter orbital period ($P_{\mathrm{t}} \sim 0\fd01$) than their original parameters shown in Fig.~\ref{0737:fig:presn}. Hence, only the helium stars represented by solid star symbols (i.e. with initial masses in the range of 4 -- 6.5~\msun) are possible progenitors of J0737--3039. This range of mass requires an initial progenitor main-sequence star of 14 -- 20~\msun.

The region of the possible pre-SN orbital period $P_{\mathrm{o}}$ limited by the two thick lines in Fig.~\ref{0737:fig:presn} does not cross the shaded area where the helium stars never fill their Roche lobes. This implies that J0737--3039 can only have been formed through Roche-lobe mass transfer from the helium star progenitor -- as opposed to the other four galactic DNSs which could be formed either via RLOF or without a mass-transfer phase (Dewi \& Pols 2003).

\begin{figure}
  	 \centerline{\includegraphics[width=84mm]{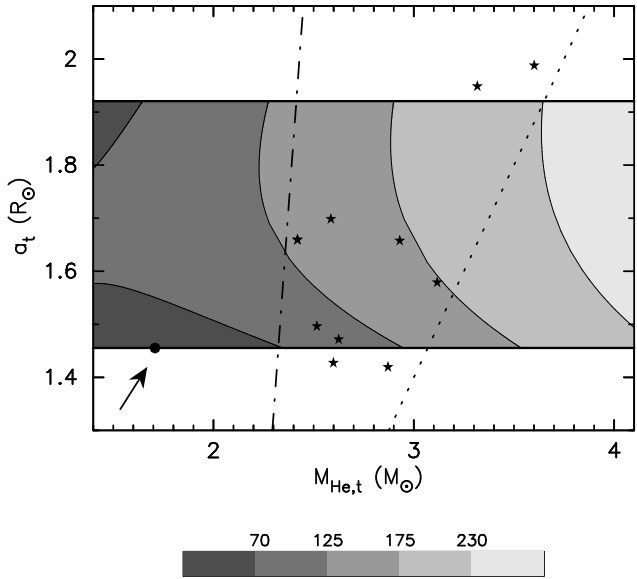}}
	 \caption[]{The possible pre-SN separation of J0737--3039 is limited between
                    the two horizontal lines. To the left of the dash-dotted line 
                    systems undergo a CE phase and may evolve to extremely short
                    orbital periods. The dotted line shows the minimum separation 
                    allowed for the a helium star to undergo a stable mass-transfer 
                    phase. To the right of this line runaway mass transfer occurs, 
                    probably leading to coalescence. The solid circle marks the pre-SN 
                    mass and separation in case of a symmetric SN explosion at the 
                    birth of the pulsar's companion. The thin lines give the magnitude 
                    of the kick velocity (the corresponding kick velocities in \kms\ 
                    are presented at the bottom of the figure).}
	 \label{0737:fig:kick}
	\end{figure}

Fig.~\ref{0737:fig:kick} depicts an enlargement of the region of allowed final HeS-NS systems from Fig.~\ref{0737:fig:presn} in an $a_{\mathrm{t}}$ vs. $M_{\mathrm{He,t}}$ diagram (the stars are the same as the allowed systems in Fig.~\ref{0737:fig:presn}). For a given combination of pre-SN helium star mass $M_{\mathrm{He,t}}$ and orbital separation $a_{\mathrm{t}}$ we calculated the kick velocity required to produce the post-SN orbital period and eccentricity. Fig.~\ref{0737:fig:kick} shows the ranges of the required kick velocities $V_{\mathrm{k}}$ limited by the curves of $V_{\mathrm{k}} =$ 70, 125, 175 and 230~\kms. The region where the systems undergo a CE phase and evolve to extremely short orbital periods is to the left of the dash-dotted line in Fig.~\ref{0737:fig:kick}. Assuming that only systems which do not go through the CE phase are the possible progenitors of J0737--3039 (i.e. the systems on the right-hand side of the line), we find that to explain the orbital parameters of J0737--3039 when it was born (i.e. $a = 1.688$~\rsun\ and $e = 0.138$; assuming that the pulsar's age equals its characteristic age), a minimum kick velocity of 70~\kms\ is required at the birth of the pulsar's companion. 

The minimum separation, below which mass-transfer phase from the helium star to a neutron star is dynamically unstable resulting in a merger of the two stars (Dewi et al. 2002), is marked by the dotted line in Fig.~\ref{0737:fig:kick}. This line gives a constraint on the maximum kick velocity, i.e. 230~\kms.

\subsection{Magnetic field and spin history of the pulsar}
\label{0737:subsec:decay}

All five double neutron star systems (DNSs) in eccentric orbits in the galactic disk share the characteristic of a combination of a weak dipole magnetic field (in the range $10^{9}$ to \xpn{2}{10}~G) and rapid spin, indicating that we are dealing here with recycled pulsars that have been spun up during an accretion phase (Smarr \& Blandford 1976; Alpar et al. 1982; Radhakrishnan \& Srinivasan 1982; Srinivasan \& van den Heuvel 1982). During such a phase with a non-degenerate donor companion the orbit will have been tidally circularized. Therefore the present eccentricity of these systems can only have been induced in the second supernova explosion when the companion collapsed to a neutron star (Srinivasan \& van den Heuvel 1982). The reduction of the magnetic field strength of the pulsar to a value around $10^{9}$ to $10^{10}$~G is thought to somehow be the result of the accretion, as there is an observed correlation between the amount of matter accreted and the field strength (Taam \& van den Heuvel 1986). A simple relation for this case was derived semi-empirically, by Shibazaki et al. (1989):
	\begin{eqnarray}
         B(t) = \frac{ B_{0} }{ 1 + \Delta M / m_{\mathrm{B}} }
        \label{0737:eq:field}
	\end{eqnarray}
where $B(t)$ and $B_{0}$ are the present and initial magnetic field strengths, respectively, $\Delta M$ is the amount of matter accreted and $m_{\mathrm{B}}$ is a constant to be derived from the observations. Francischelli et al. (2002) found $m_{\mathrm{B}} =$ \xpn{1.25}{-5}~\msun\ to agree well with the observed data for Low Mass X-ray Binaries. Adopting this value and a duration of the Roche-lobe overflow phase of the helium star of \xpn{2}{4}~yrs (this is about the average value for helium stars in the mass range 4.0 to 6.3~\msun\ (Dewi et al. 2002; Dewi \& Pols 2003)), with accretion at an Eddington rate for helium (\xpn{3}{-8}~\rate) one finds that $B(t)$ can have decreased by a factor of about 50 during this phase alone. Furthermore, a 15~\msun\ Be star resides $10^{7}$~yrs on the main-sequence. Be/X-ray binaries tend to be transients (Rappaport \& van den Heuvel 1982; van den Heuvel \& Rappaport 1987; Zi{\'o}{\l}kowski 2002) and are "on" as an X-ray source only part of the time, during which they often reach a luminosity close to the Eddington limit. Assuming them to be "on" only one per cent of the time and then to reach the Eddington limit, the neutron star will during the Be/X-ray binary phase have accreted about \xpn{1.5}{-3}~\msun. With eq.~(\ref{0737:eq:field}) this is already sufficient to reduce its magnetic field by a factor of about 100. So, adding the amount of matter accreted during the Be/X-ray binary phase and the RLOF of the helium star, one finds with eq.~(\ref{0737:eq:field}) that these phases together provide enough accretion to reduce the magnetic field by a factor of about 200. 

Magnetic fields of young pulsars on average typically have a dipole strength a few times $10^{12}$~G, and after reduction by this factor will typically be around $10^{10}$~G, as  observed in the five DNSs. As to the spin-up of the "old" neutron star: during the Roche-lobe overflow from the helium star accretion to the neutron star will have proceeded through a disk. The amount of angular momentum accreted during this phase can be calculated as follows. The mass transfer rate during Roche-lobe overflow is highly super-Eddington, in the range \xpn{1-2}{-4}~\rate\ for helium stars in the mass range 4.0 -- 6.3~\msun\ (Dewi et al. 2002; Dewi \& Pols 2003). Dewi et al. (2002) and Dewi \& Pols (2003) in their calculations of the evolution of the systems during this phase assumed the accretion rate onto the neutron star to be the Eddington rate (\xpn{3}{-8}~\rate) and the super-Eddington fraction of the transferred matter to be blown out of the system by the radiation pressure, carrying with it the specific angular momentum equal to the specific orbital angular momentum of the neutron star. Following this assumption of accretion at the Eddington rate -- as radiation pressure will prevent accretion at a higher rate -- and using the dipole strength of the magnetic field of J0737--3039 of \xpn{6.9}{9}~G, one can calulate the value of the Alfven-radius $R_{\mathrm{a}}$, where the matter becomes coupled to the magnetic field and starts to co-rotate with the neutron star magnetosphere. $R_{\mathrm{a}}$ is given by (e.g. see Bhattacharya \& van den Heuvel 1991)): 
	\begin{eqnarray}
          R_{\mathrm{a}} = \frac{(B_{9} \, R_{6}^{3})^{4/7}}{(\dot{M}/{\dot{M}}_{\mathrm{E,H}})^{2/7} \,\, (M/\mathrm{M_{\sun}})^{1/7}} . 32.9 \, \mathrm{km}
        \label{0737:eq:alfven}
	\end{eqnarray}
where $B_{9}$ is the dipole field strength in units of $10^{9}$~G, $R_{6}$ is the neutron star radius in units of $10^{6}$ cm, $\dot{M}$ and ${\dot{M}}_{\mathrm{E,H}}$ are the accretion rate and the Eddington accretion rate for hydrogen, respectively and $M$ is the mass of the neutron star. With the dipole field strength of \xpn{6.9}{9}~G of PSR J0737--3039, the Eddington rate for helium accretion $\dot{M} = {\dot{M}}_{\mathrm{E,He}} =$ \xpn{3}{-8}~\rate\ and a $M$ = 1.3~\msun\ one obtains $R_{\mathrm{a}}$ = 78.4~km. 

The rate at which angular momentum is accreted by the neutron star is then:
	\begin{eqnarray}
          \dot{J} = \dot{M} \, \Omega_{\mathrm{k}}(r=R_{\mathrm{a}}) \, R_{\mathrm{a}}^2
        \label{0737:eq:jdot}
	\end{eqnarray}
where $\Omega_{\mathrm{k}}(r)$ is the keplerian angular velocity of matter in the disk at distance $r$ from the center of the neutron star. With $\dot{M} = {\dot{M}}_{\mathrm{E,He}}$, $M = 1.3$~\msun\ one finds $\dot{J} = 1.1 \times 10^{-1} \mathrm{M_{\sun} \, km^2 \, s^{-1} \, yr^{-1}}$. The present spin angular momentum of the neutron star is 
	\begin{eqnarray}
          J = M (2 \pi/P) \, R^2 r_{\mathrm{g}}^2
        \label{0737:eq:spin}
	\end{eqnarray}
where $R$ and $P$ are the radius and spin period of the neutron star and $r_{\mathrm{g}}$ is its radius of gyration, respectively. The value of $r_{\mathrm{g}}^{2}$ is typically about 0.075 for a polytropic index  $n = 3$ (e.g. Lattimer \& Prakash 2001). With this value and $R = 10$~km and $P$ = 22 ms one obtains $J = 2.8 \times 10^{3} \,\, \mathrm{M_{\sun} \, km^2 \, s^{-1}}$, which together with the above mentioned rate of accretion of angular momentum results in a required timescale for the spin-up of about \xpn{2.5}{4} years.

The duration of the phase of Roche-lobe overflow in the helium star plus neutron star binaries with helium star masses between 4.0 and 6.3~\msun\ ranges from 2.4 to \xpn{0.7}{4} years, which is of the same order as the value of \xpn{2.5}{4} years required for the spin-up. It thus appears that the magnetic field strength and spin rate of PSR J0737--3039 fit quite well with evolutionary model presented here in which the immediate progenitor system was a very close helium star plus neutron star binary which evolved through a phase of Roche lobe overflow with a duration of order $10^4$ years. As mentioned above, the progenitor of this system was so close that it cannot have avoided the phase of mass transfer by Roche-lobe overflow from the helium star (Dewi \& Pols 2003), so here this phase must absolutely have taken place in the history of PSR J0737--3039. This may be the reason why the pulsar has the shortest spin period of all. The progenitors of the other systems may, in principle, have avoided RLOF, although it can certainly not be excluded that also in (some of) them Roche-lobe overflow from the helium star may have taken place.             

\section{Discussions and Conclusions}
\label{0737:sec:conclusions}

We have examined the formation of J0737--3039. We find that this system must have gone through a mass-transfer phase from a helium star with the initial mass in the range of 4 -- 6.3~\msun. We also find that an asymmetric SN explosion with a kick velocity of 70 -- 230~\kms\ is necessary to explain the observed orbital parameters. However, these constraints may still be subject to some uncertainties. Whether the neutron stars in the systems that go through a CE phase (the open stars in Fig.~\ref{0737:fig:presn}) complete the spiral-in in the envelope of the helium star depends on the competition between the spiral-in timescale and the time left before the explosion of the helium star (Dewi \& Pols 2003). If the SN explosion in systems with open star symbols in Fig.~\ref{0737:fig:presn} occurs before the neutron star completes the spiral-in process, a lower kick velocity or even a symmetric SN explosion (shown as the solid circle) would then be possible. Furthermore, the constraints on the maximum pre-SN mass and the maximum kick velocity depend somewhat on the assumed wind mass-loss rate and the loss of angular momentum during the RLOF. As the stability of a mass-transfer phase depends on the mass ratio of the system and the period at the onset of RLOF, applying a higher wind mass-loss rate results in a lower mass helium star in wider orbit, and hence, more stable mass transfer than adopting wind mass loss with a lower rate.

\section*{Acknowledgements}
This work was sponsored by NWO Spinoza Grant 08-0 to E.~P.~J. van den Heuvel. We thank Dick Manchester for releasing the observational data of the system prior to publication, Onno Pols and Andrea Possenti for suggestions and discussions, and Hans Ritter for his referee comments.

\label{lastpage}

\end{document}